\documentstyle[12pt]{article}

\begin{document}

\author{C. Bizdadea\thanks{%
e-mail addresses: bizdadea@central.ucv.ro and bizdadea@hotmail.com}, L.
Saliu, S. O. Saliu\thanks{%
e-mail addresses: osaliu@central.ucv.ro and odile\_saliu@hotmail.com} \\
Department of Physics, University of Craiova\\
13 A. I. Cuza Str., Craiova RO-1100, Romania}
\title{Chapline-Manton interaction vertices and Hamiltonian BRST cohomology }
\maketitle

\begin{abstract}
Consistent interactions between Yang-Mills gauge fields and an abelian
2-form are investigated by using a Hamiltonian cohomological procedure. It
is shown that the deformation of the BRST charge and the BRST-invariant
Hamiltonian of the uncoupled model generates the Yang-Mills Chern-Simons
interaction term. The resulting interactions deform both the gauge
transformations and their algebra, but not the reducibility relations.

\end{abstract}

\section{Introduction}

The problem of consistent interactions that can be introduced among fields
with gauge freedom in such a way to preserve the number of gauge symmetries 
\cite{1}--\cite{4} has been reformulated as a deformation problem of the
master equation \cite{5} in the context of the antifield-BRST formalism \cite
{6}--\cite{10}. This technique has been applied to Chern-Simons models \cite
{5}, Yang-Mills theories \cite{11} and two-form gauge fields \cite{12}.
Thus, the antifield BRST method was proved to be an elegant tool for
analyzing the problem of consistent interactions.

In this paper we study another interesting interaction, namely, the
consistent interaction between the Yang-Mills vector potential and an
abelian two-form, but from the Hamiltonian BRST point of view \cite{10}, 
\cite{13}--\cite{17}. Our procedure will lead to combined
Yang-Mills-two-form system coupled through a Yang-Mills Chern-Simons term
(the Chapline-Manton model) \cite{18}--\cite{21}. Chern-Simons couplings of
a two-form to Yang-Mills theory play a major role in the Green-Schwarz
anomaly cancellation mechanism \cite{22}, and hence are useful in string
theory \cite{23}. On the other hand, the Hamiltonian BRST approach appears
to be a more natural setting for implementing the BRST symmetry in quantum
mechanics \cite{10} (chapter 14), as well as for establishing a proper
connection with canonical quantization formalisms, like for instance the
reduced phase-space or Dirac quantization procedures \cite{24}. To our
knowledge, the Hamiltonian approach to consistent interactions among fields
with gauge freedom has not been investigated until now, so our paper
establishes a new result.

The strategy to be developed is the following. Initially, we begin with the
``free'' model describing pure Yang-Mills theory and a free abelian two-form
and determine its main Hamiltonian BRST ingredients, namely, the BRST charge
and BRST-invariant Hamiltonian. The BRST symmetry of the uncoupled theory, $%
s $, can be conveniently written like the sum between the Koszul-Tate
differential and the exterior derivative along the gauge orbits, $s=\delta
+\gamma $. Subsequently, we pass to the deformation procedure along the
lines of a cohomological approach. Thus, we start by writing down the
general equations representing the core of the Hamiltonian deformation
procedure, which describe the deformation of the BRST charge, respectively,
of the BRST-invariant Hamiltonian of the ``free'' theory. Then, we proceed
to solve the main equations in relation with the model under study taking
into account the BRST cohomology of the ``free'' theory. In this way, we
reach the BRST charge and BRST-invariant Hamiltonian underlying the deformed
model. Further, we identify the Hamiltonian system behind the deformation
procedure by analyzing its first-class constraints, first-class Hamiltonian
and the accompanying gauge algebra. It turns out that the resulting system
is nothing but the Yang-Mills theory coupled to the 2-form through the
Yang-Mills Chern-Simons interaction term, also known as the Chapline-Manton
model.

\section{Hamiltonian BRST symmetry for the uncoupled theory}

In this section we derive the Hamiltonian BRST symmetry for the ``free''
theory. In this respect, we begin with a Lagrangian action equal with the
sum between the actions of Yang-Mills theory and a free 2-form 
\begin{equation}  \label{1}
S_0^L\left[ A_\mu ^a,B_{\mu \nu }\right] =\int d^Dx\left( -\frac 14F_{\mu
\nu }^aF_a^{\mu \nu }-\frac 1{12}F_{\mu \nu \rho }F^{\mu \nu \rho }\right) ,
\end{equation}
where 
\begin{equation}  \label{2}
F_{\mu \nu }^a=\partial _\mu A_\nu ^a-\partial _\nu A_\mu
^a-f_{\;\;bc}^aA_\mu ^bA_\nu ^c,
\end{equation}
\begin{equation}  \label{3}
F_{\mu \nu \rho }=\partial _\mu B_{\nu \rho }+\partial _\rho B_{\mu \nu
}+\partial _\nu B_{\rho \mu }\equiv \partial _{\left[ \mu \right. }B_{\left.
\nu \rho \right] }.
\end{equation}
The canonical analysis of action (\ref{1}) gives the first-class constraints 
\begin{equation}  \label{4}
G_a^{(1)}\equiv \pi _a^0\approx 0,\;G_i^{(1)}\equiv \pi _{0i}\approx 0,
\end{equation}
\begin{equation}  \label{5}
G_a^{(2)}\equiv -\left( \partial _j\pi _a^j-f_{\;\;ac}^b\pi _b^jA_j^c\right)
\approx 0,\;G_i^{(2)}\equiv -2\partial ^j\pi _{ji}\approx 0,
\end{equation}
and the first-class Hamiltonian 
\begin{eqnarray}  \label{6}
& &H_0=\int d^{D-1}x\left( \frac 12\pi _{aj}\pi _j^a+\frac
14F_{ij}^aF_a^{ij}+A_0^aG_a^{(2)}-\right.  \nonumber \\
& &\left. \pi _{ij}\pi ^{ij}+\frac
1{12}F_{ijk}F^{ijk}+B^{0i}G_i^{(2)}\right) .
\end{eqnarray}
In (\ref{4}--\ref{6}), $\pi _a^\mu $ and $\pi _{\mu \nu }$ denote the
canonical momenta of $A_\mu ^a$, respectively, $B^{\mu \nu }$. The gauge
algebra of the uncoupled model reads as 
\begin{equation}  \label{7}
\left[ G_a^{(1)},G_b^{(1)}\right] =0,\;\left[ G_a^{(1)},G_b^{(2)}\right]
=0,\;\left[ G_a^{(2)},G_b^{(2)}\right] =f_{\;\;ab}^cG_c^{(2)},
\end{equation}
\begin{equation}  \label{8}
\left[ G_i^{(1)},G_j^{(1)}\right] =0,\;\left[ G_i^{(1)},G_j^{(2)}\right]
=0,\;\left[ G_i^{(2)},G_j^{(2)}\right] =0,
\end{equation}
\begin{equation}  \label{9}
\left[ G_a^{(1)},G_i^{(1)}\right] =0,\;\left[ G_a^{(1)},G_i^{(2)}\right]
=0,\;\left[ G_a^{(2)},G_i^{(1)}\right] =0,\;\left[
G_a^{(2)},G_i^{(2)}\right] =0,
\end{equation}
\begin{equation}  \label{10}
\left[ H_0,G_a^{(1)}\right] =G_a^{(2)},\;\left[ H_0,G_a^{(2)}\right]
=-f_{\;\;ab}^cA_0^bG_c^{(2)},
\end{equation}
\begin{equation}  \label{11}
\left[ H_0,G_i^{(1)}\right] =G_i^{(2)},\;\left[ H_0,G_i^{(2)}\right] =0.
\end{equation}
In addition, the constraint functions $G_i^{(2)}$ are first-stage reducible,
i.e., 
\begin{equation}  \label{12}
\partial ^iG_i^{(2)}=0.
\end{equation}
On account of (\ref{7}--\ref{12}), the BRST charge and BRST-invariant
Hamiltonian of the uncoupled theory are given by 
\begin{eqnarray}  \label{13}
& &\Omega _0=\int d^{D-1}x\left( G_a^{(1)}\eta _1^a+G_a^{(2)}\eta _2^a+\frac
12f_{\;\;bc}^a{\cal P}_{2a}\eta _2^b\eta _2^c+\right.  \nonumber \\
& &\left. G_i^{(1)}\eta _1^i+G_i^{(2)}\eta _2^i+\eta \partial ^i{\cal P}%
_{2i}\right) ,
\end{eqnarray}
\begin{equation}  \label{14}
H_B=H_0+\int d^{D-1}x\left( \left( \eta _1^a-f_{\;\;bc}^a\eta
_2^bA_0^c\right) {\cal P}_{2a}+\eta _1^i{\cal P}_{2i}\right) .
\end{equation}
In the above, $\eta _1^a$, $\eta _2^a$, $\eta _1^i$ and $\eta _2^i$ stand
for the fermionic ghost number one Hamiltonian ghosts, $\eta $ denotes the
bosonic ghost number two ghost for ghost, while the ${\cal P}$'s represent
their corresponding canonical momenta (antighosts). The ghost number is
defined like the difference between the pure ghost number ($pgh$) and the
antighost number ($antigh$), with 
\begin{equation}  \label{15}
pgh\left( z^A\right) =0,\;pgh\left( \eta ^\Gamma \right) =1,\;pgh\left( \eta
\right) =2,pgh\left( {\cal P}_\Gamma \right) =0,\;pgh\left( {\cal P}\right)
=0,
\end{equation}
\begin{equation}  \label{16}
antigh\left( z^A\right) =0,\;antigh\left( \eta ^\Gamma \right)
=0,\;antigh\left( \eta \right) =0,
\end{equation}
\begin{equation}  \label{17}
antigh\left( {\cal P}_\Gamma \right) =1,\;antigh\left( {\cal P}\right) =2,
\end{equation}
where 
\begin{equation}  \label{18}
z^A=\left( A_\mu ^a,B^{\mu \nu },\pi _a^\mu ,\pi _{\mu \nu }\right) ,\eta
^\Gamma =\left( \eta _1^a,\eta _2^a,\eta _1^i,\eta _2^i\right) ,{\cal P}%
_\Gamma =\left( {\cal P}_{1a},{\cal P}_{2a},{\cal P}_{1i},{\cal P}%
_{2i}\right) .
\end{equation}
The BRST differential $s\bullet =\left[ \bullet ,\Omega _0\right] $ of the
uncoupled theory splits as 
\begin{equation}  \label{19}
s=\delta +\gamma ,
\end{equation}
where $\delta $ is the Koszul-Tate differential, and $\gamma $ represents
the exterior longitudinal derivative along the gauge orbits. These operators
act like 
\begin{equation}  \label{20}
\delta z^A=0,\;\delta \eta ^\Gamma =0,\;\delta \eta =0,
\end{equation}
\begin{equation}  \label{21}
\delta {\cal P}_{1a}=-\pi _a^0,\;\delta {\cal P}_{2a}=\partial _j\pi
_a^j-f_{\;\;ac}^b\pi _b^jA_j^c,
\end{equation}
\begin{equation}  \label{22}
\delta {\cal P}_{1i}=-\pi _{0i},\;\delta {\cal P}_{2i}=2\partial ^j\pi
_{ji},\;\delta {\cal P}=-\partial ^i{\cal P}_{2i},
\end{equation}
\begin{equation}  \label{23}
\gamma A_0^a=\eta _1^a,\;\gamma A_i^a=\partial _i\eta _2^a+f_{\;\;bc}^a\eta
_2^bA_i^c,\;\gamma B^{0i}=\eta _1^i,\;\gamma B^{ij}=\partial ^{\left[
i\right. }\eta _2^{\left. j\right] },
\end{equation}
\begin{equation}  \label{24}
\gamma \pi _a^0=0,\;\gamma \pi _a^i=f_{\;\;ac}^b\pi _b^i\eta _2^c,\;\gamma
\pi _{0i}=0,\;\gamma \pi _{ij}=0,
\end{equation}
\begin{equation}  \label{25}
\gamma \eta _1^a=0,\;\gamma \eta _2^a=-\frac 12f_{\;\;bc}^a\eta _2^b\eta
_2^c,\;\gamma \eta _1^i=0,\;\gamma \eta _2^i=\partial ^i\eta ,\;\gamma \eta
=0,
\end{equation}
\begin{equation}  \label{26}
\gamma {\cal P}_{1a}=0,\;\gamma {\cal P}_{2a}=f_{\;\;ab}^c{\cal P}_{2c}\eta
_2^b,\;\gamma {\cal P}_{1i}=0,\;\gamma {\cal P}_{2i}=0,\;\gamma {\cal P}=0.
\end{equation}
Formulas (\ref{20}--\ref{26}) will be employed in the next section in the
framework of the deformation procedure.

\section{Deformation of the ``free'' theory}

In this section we deform the uncoupled model discussed above in the
framework of the Hamiltonian BRST formalism. First, we write down the
general equations underlying the deformation of the BRST charge and
BRST-invariant Hamiltonian. Second, we solve these equations with respect to
the model under study by using the cohomological technique. Finally, we
identify the new gauge theory, which turns out to be nothing but the
Chapline-Manton model.

\subsection{Hamiltonian deformation problem}

It is well-known that the solution to the master equation captures all the
information on a given gauge theory at the level of the antifield BRST
formalism. The gauge-fixed dynamics is generated by the gauge-fixed action,
which is obtained from the solution to the master equation by using a
certain gauge-fixing fermion. Moreover, it has been shown that the
deformation of the solution to the master equation generates consistent
interactions among fields with gauge freedom \cite{5}. At the Hamiltonian
level, the BRST charge $\Omega _0$ contains all the information on the
structure of a first-class system. In this sense, the BRST charge plays a
role similar to that of the solution to the master equation. However, in
order to stipulate the correct dynamics, one needs a Hamiltonian, which is
nothing but the gauge-fixed Hamiltonian $H_K=H_B+\left[ K,\Omega _0\right] $%
, where $H_B$ stands for the BRST-invariant Hamiltonian and $K$ is the
gauge-fixing fermion. Thus, we can conclude that the problem of deforming
the master equation induces at the Hamiltonian level the deformation of the
equation $\left[ \Omega _0,\Omega _0\right] =0$, as well as of the
BRST-invariant Hamiltonian of the ``free'' theory.

The Lagrangian deformation implies that the BRST charge of the uncoupled
theory is deformed as 
\begin{eqnarray}  \label{27}
& &\Omega _0\rightarrow \Omega =\Omega _0+g\int d^{D-1}\omega _1+g^2\int
d^{D-1}\omega _2+O\left( g^3\right) =  \nonumber \\
& &\Omega _0+g\Omega _1+g^2\Omega _2+O\left( g^3\right) ,
\end{eqnarray}
where $\Omega $ should satisfy the equation 
\begin{equation}  \label{28}
\left[ \Omega ,\Omega \right] =0.
\end{equation}
Equation (\ref{28}) splits accordingly the deformation parameter as 
\begin{equation}  \label{29}
\left[ \Omega _0,\Omega _0\right] =0,
\end{equation}
\begin{equation}  \label{30}
2\left[ \Omega _0,\Omega _1\right] =0,
\end{equation}
\begin{equation}  \label{31}
2\left[ \Omega _0,\Omega _2\right] +\left[ \Omega _1,\Omega _1\right] =0,
\end{equation}
\[
\vdots 
\]
Obviously, equation (\ref{29}) is automatically satisfied. From the
remaining equations we deduce the pieces $\left( \Omega _k\right) _{k>0}$ on
account of the ``free'' BRST differential. With the deformed BRST charge at
hand, we then deform the BRST-invariant Hamiltonian of the ``free'' theory 
\begin{eqnarray}  \label{32}
& &H_B\rightarrow \tilde H_B= H_B+g\int d^{D-1}h_1+g^2\int
d^{D-1}h_2+O\left( g^3\right) =  \nonumber \\
& &H_B+gH_1+g^2H_2+O\left( g^3\right) ,
\end{eqnarray}
and require that 
\begin{equation}  \label{33}
\left[ \tilde H_B,\Omega \right] =0.
\end{equation}
Like in the previous case, equation (\ref{33}) can be decomposed accordingly
the deformation parameter like 
\begin{equation}  \label{34}
\left[ H_B,\Omega _0\right] =0,
\end{equation}
\begin{equation}  \label{35}
\left[ H_B,\Omega _1\right] +\left[ H_1,\Omega _0\right] =0,
\end{equation}
\begin{equation}  \label{36}
\left[ H_B,\Omega _2\right] +\left[ H_1,\Omega _1\right] +\left[ H_2,\Omega
_0\right] =0,
\end{equation}
\[
\vdots 
\]
Clearly, equation (\ref{34}) is again fulfilled, while from the other
equations one can determine the components $\left( H_k\right) _{k\geq 1}$
relying on the BRST symmetry of the ``free'' model.

\subsection{Deformation of BRST charge}

Here, we solve the equations (\ref{30}--\ref{31}) in the context of the
uncoupled model under discussion taking into account that the ``free'' BRST
differential splits as in (\ref{19}). Equation (\ref{30}) holds if and only
if $\omega _{1}$ is a $s$-co-cycle modulo $\tilde{d}=dx^{i}\partial _{i}$,
i.e., 
\begin{equation}
s\omega _{1}=\partial _{k}j^{k},  \label{37}
\end{equation}
for some $j^{k}$. In order to solve equation (\ref{37}) we expand $\omega
_{1}$ according to the antighost number 
\begin{equation}
\omega _{1}=\stackrel{(0)}{\omega }_{1}+\stackrel{(1)}{\omega }_{1}+\cdots 
\stackrel{(J)}{\omega }_{1},\;antigh\left( \stackrel{(I)}{\omega }%
_{1}\right) =I,  \label{38}
\end{equation}
where the last term in (\ref{38}) can be assumed to be annihilated by $%
\gamma $. Since $antigh\left( \stackrel{(J)}{\omega }_{1}\right) =J$ and $%
gh\left( \stackrel{(J)}{\omega }_{1}\right) =1$, it follows that $pgh\left( 
\stackrel{(J)}{\omega }_{1}\right) =J+1$. On the other hand, we have that 
\begin{equation}
\rho =\frac{1}{3}f_{abc}\eta _{2}^{a}\eta _{2}^{b}\eta _{2}^{c},  \label{39}
\end{equation}
and the ghost for ghost $\eta $ are $\gamma $-invariant, hence we can
represent $\stackrel{(J)}{\omega }_{1}$ as 
\begin{equation}
\stackrel{(J)}{\omega }_{1}=\mu _{J}\sum\limits_{N,M}\left( \rho \right)
^{N}\left( \eta \right) ^{M},  \label{38a}
\end{equation}
where $N$, $M$ are some nonnegative integers with $3N+2M=J+1$. With this
choice, it is easy to check that the $\gamma $-invariant coefficient $\mu
_{J}$ belongs to $H_{J}\left( \delta |\tilde{d}\right) $. Using the results
from \cite{25} adapted to the Hamiltonian treatment, it follows that $%
H_{J}\left( \delta |\tilde{d}\right) =0$ for $J>2$ in the case of our
uncoupled model. This means that the last term in (\ref{38}) corresponds to $%
J=2$, which then leads to $3N+2M=3$. As a consequence, we have that $N=1$, $%
M=0$ (the ghost for ghost brings no contribution), such that (\ref{38})
takes the form 
\begin{equation}
\omega _{1}=\stackrel{(0)}{\omega }_{1}+\stackrel{(1)}{\omega }_{1}+%
\stackrel{(2)}{\omega }_{1},  \label{41}
\end{equation}
where 
\begin{equation}
\stackrel{(2)}{\omega }_{1}=\frac{1}{3}\mu _{2}f_{abc}\eta _{2}^{a}\eta
_{2}^{b}\eta _{2}^{c},  \label{42}
\end{equation}
and $\mu _{2}$ from $H_{2}\left( \delta |\tilde{d}\right) $, therefore
solution to the equation 
\begin{equation}
\delta \mu _{2}+\partial _{k}v^{k}=0,  \label{43}
\end{equation}
for some $v^{k}$. From the last relation in (\ref{22}) we find that $\mu
_{2}={\cal P}$, so 
\begin{equation}
\stackrel{(2)}{\omega }_{1}=\frac{1}{3}f_{abc}{\cal P}\eta _{2}^{a}\eta
_{2}^{b}\eta _{2}^{c}.  \label{44}
\end{equation}
At antighost number one, equation (\ref{37}) takes the form 
\begin{equation}
\delta \stackrel{(2)}{\omega }_{1}+\gamma \stackrel{(1)}{\omega }%
_{1}=\partial _{k}u^{k}.  \label{45}
\end{equation}
Starting from 
\begin{equation}
\delta \stackrel{(2)}{\omega }_{1}=\frac{1}{3}f_{abc}\left( \partial ^{i}%
{\cal P}_{2i}\right) \eta _{2}^{a}\eta _{2}^{b}\eta _{2}^{c},  \label{46}
\end{equation}
we deduce 
\begin{equation}
\stackrel{(1)}{\omega }_{1}=-f_{abc}{\cal P}_{2i}\eta _{2}^{a}\eta
_{2}^{b}A^{ci},  \label{47}
\end{equation}
such that 
\begin{equation}
\delta \stackrel{(2)}{\omega }_{1}+\gamma \stackrel{(1)}{\omega }%
_{1}=\partial ^{i}\left( \frac{1}{3}f_{abc}{\cal P}_{2i}\eta _{2}^{a}\eta
_{2}^{b}\eta _{2}^{c}\right) .  \label{48}
\end{equation}
At antighost number zero, equation (\ref{37}) is given by 
\begin{equation}
\delta \stackrel{(1)}{\omega }_{1}+\gamma \stackrel{(0)}{\omega }%
_{1}=\partial _{k}w^{k}.  \label{49}
\end{equation}
On account of (\ref{47}), it results that 
\begin{equation}
\delta \stackrel{(1)}{\omega }_{1}=2f_{abc}\eta _{2}^{a}\eta
_{2}^{b}A_{i}^{c}\partial _{j}\pi ^{ji},  \label{50}
\end{equation}
which further leads to 
\begin{equation}
\stackrel{(0)}{\omega }_{1}=4\pi ^{ji}\left( \partial _{j}A_{ai}\right) \eta
_{2}^{a},  \label{51}
\end{equation}
such that 
\begin{equation}
\delta \stackrel{(1)}{\omega }_{1}+\gamma \stackrel{(0)}{\omega }%
_{1}=\partial _{j}\left( 2f_{abc}\eta _{2}^{a}\eta _{2}^{b}A_{i}^{c}\pi
^{ji}\right) .  \label{52}
\end{equation}
Thus, we have generated the first-order deformation of the BRST charge under
the form 
\begin{equation}
\Omega _{1}=\int d^{D-1}x\left( 4\pi ^{ji}\left( \partial _{j}A_{ai}\right)
\eta _{2}^{a}-f_{abc}{\cal P}_{2i}\eta _{2}^{a}\eta _{2}^{b}A^{ci}+\frac{1}{3%
}f_{abc}{\cal P}\eta _{2}^{a}\eta _{2}^{b}\eta _{2}^{c}\right) .  \label{53}
\end{equation}
The deformation is consistent also to order $g^{2}$ if and only if $\left[
\Omega _{1},\Omega _{1}\right] $ is $s$-exact (see (\ref{31})). It is easy
to see that $\left[ \Omega _{1},\Omega _{1}\right] =0$, so $\Omega _{2}=0$.
The higher-order equations are then satisfied with $\Omega _{3}=\Omega
_{4}=\cdots =0$. In this way, we inferred that $\Omega =\Omega _{0}+g\Omega
_{1}$ is a complete solution for the equation (\ref{28}) that describes the
deformation of the BRST charge.

\subsection{Deformation of BRST-invariant Hamiltonian}

Next we pass to determine the deformation of the BRST-invariant Hamiltonian (%
\ref{14}). Initially, we compute $H_{1}$ as solution to the equation (\ref
{35}). Simple calculations lead to the expression of the first term in (\ref
{35}) of the type 
\begin{eqnarray}
&&\left[ H_{B},\Omega _{1}\right] =\int d^{D-1}x\left( f_{abc}{\cal P}%
_{2i}\eta _{2}^{a}\eta _{2}^{b}\left( \pi ^{ci}-\partial
^{i}A_{0}^{c}-f_{\;\;de}^{c}A_{0}^{d}A^{ei}\right) -\right.   \nonumber
\label{54} \\
&&4\left( \pi _{a}^{i}-f_{abc}A_{0}^{b}A^{ci}\right) \partial ^{j}\left( \pi
_{ji}\eta _{2}^{a}\right) -2\left( \partial _{i}F^{ijk}\right) \left(
\partial _{j}A_{ak}\right) \eta _{2}^{a}-  \nonumber \\
&&\left. \left( \eta _{1}^{c}-f_{\;\;de}^{c}A_{0}^{e}\eta _{2}^{d}\right)
\left( f_{abc}{\cal P}\eta _{2}^{a}\eta _{2}^{b}+4\pi ^{ji}\partial
_{j}A_{ci}+2f_{abc}{\cal P}_{2i}A^{ai}\eta _{2}^{b}\right) \right) =  
\nonumber \\
&&\int d^{D-1}x\lambda .
\end{eqnarray}
In consequence, (\ref{35}) gives 
\begin{equation}
sh_{1}+\lambda =\partial _{k}\alpha ^{k},  \label{55}
\end{equation}
for some $\alpha ^{k}$. Then, we further obtain 
\begin{eqnarray}
&&h_{1}=4\left( A_{a}^{0}\partial _{i}A_{j}^{a}-\pi _{ai}A_{j}^{a}\right)
\pi ^{ij}-\frac{1}{3}F^{ijk}\left(
f_{abc}A_{i}^{a}A_{j}^{b}A_{k}^{c}+A_{\left[ i\right. }^{a}F_{a\left.
jk\right] }\right) +  \nonumber  \label{56} \\
&&2\left( \pi _{a}^{i}+f_{abc}A_{0}^{c}A^{ib}\right) \eta _{2}^{a}{\cal P}%
_{2i}+f_{abc}A_{0}^{c}{\cal P}\eta _{2}^{a}\eta _{2}^{b},
\end{eqnarray}
such that 
\begin{equation}
sh_{1}+\lambda =\partial _{k}\left( \left( f_{abc}A_{0}^{c}{\cal P}%
_{2}^{k}\eta _{2}^{b}-2F^{ijk}\partial _{i}A_{aj}\right) \eta
_{2}^{a}\right) .  \label{57}
\end{equation}
With $h_{1}$ at hand, we pass to solve equation (\ref{36}). The first term
in (\ref{36}) vanishes ($\Omega _{2}=0$), while the second term is given by 
\begin{eqnarray}
&&\left[ H_{1},\Omega _{1}\right] =\int d^{D-1}x\left( 4\left( f_{abc}{\cal P%
}_{2}^{i}\eta _{2}^{a}\eta _{2}^{b}-4\partial _{j}\left( \pi ^{_{ji}}\eta
_{2c}\right) A^{ck}\pi _{ik}\right) +\right.   \nonumber  \label{58} \\
&&\left. 2\left( \partial ^{i}\left(
f_{abc}A_{i}^{a}A_{j}^{b}A_{k}^{c}+A_{\left[ i\right. }^{a}F_{a\left.
jk\right] }\right) \right) \left( \partial ^{\left[ j\right. }A_{d}^{\left.
k\right] }\right) \eta _{2}^{d}\right) =\int d^{D-1}x\nu .
\end{eqnarray}
Therefore, equation (\ref{36}) implies 
\begin{equation}
sh_{2}+\nu =\partial ^{i}\beta _{i}.  \label{59}
\end{equation}
The solution to (\ref{59}) reads as 
\begin{equation}
h_{2}=-8A^{ka}A_{ja}\pi _{ik}\pi ^{ij}+\frac{1}{3}\left(
f_{abc}A_{i}^{a}A_{j}^{b}A_{k}^{c}+A_{\left[ i\right. }^{a}F_{a\left.
jk\right] }\right) ^{2}+8A_{a}^{j}\pi _{ij}\eta _{2}^{a}{\cal P}_{2}^{i},
\label{60}
\end{equation}
so we find that 
\begin{equation}
sh_{2}+\nu =\partial ^{i}\left( 2\left(
f_{abc}A_{i}^{a}A_{j}^{b}A_{k}^{c}+A_{\left[ i\right. }^{a}F_{a\left.
jk\right] }\right) \left( \partial ^{\left[ j\right. }A_{d}^{\left. k\right]
}\right) \eta _{2}^{d}\right) .  \label{61}
\end{equation}
In this manner, we inferred also the order $g^{2}$ deformation of the
BRST-invariant Hamiltonian. The equation describing the order $g^{3}$ 
deformation is
clearly satisfied for $h_{3}=0$ because all the terms that do not involve $%
h_{3}$ vanish. The higher-order deformation equations are then fulfilled for 
$h_{4}=h_{5}=\cdots =0$. In conclusion, $\tilde{H}_{B}=H_{B}+g\int
d^{D-1}xh_{1}+g^{2}\int d^{D-1}xh_{2}$, with $h_{1}$ and $h_{2}$ expressed
by (\ref{56}), respectively, (\ref{60}), is solution to the deformation
problem of the BRST-invariant Hamiltonian.

\subsection{Identification of the new gauge theory}

Putting together the results deduced in the previous two subsections, we
remark that the complete solutions to the deformation problems related to
the BRST charge and BRST-invariant Hamiltonian are pictured by 
\begin{eqnarray}
&&\Omega =\int d^{D-1}x\left( -\left( \partial _{j}\pi
_{a}^{j}-f_{\;\;ac}^{b}\pi _{b}^{j}A_{j}^{c}-4g\pi ^{ij}\partial
_{i}A_{aj}\right) \eta _{2}^{a}+\right.   \nonumber  \label{62} \\
&&\pi _{a}^{0}\eta _{1}^{a}+\pi _{0i}\eta _{1}^{i}+\eta \partial ^{i}{\cal P}%
_{2i}-2\left( \partial ^{j}\pi _{ji}\right) \eta _{2}^{i}+  \nonumber \\
&&\left. \frac{1}{2}f_{\;\;ab}^{c}\left( {\cal P}_{2c}+2gA_{c}^{i}{\cal P}%
_{2i}\right) \eta _{2}^{a}\eta _{2}^{b}+\frac{g}{3}f_{abc}{\cal P}\eta
_{2}^{a}\eta _{2}^{b}\eta _{2}^{c}\right) ,
\end{eqnarray}
respectively, 
\begin{eqnarray}
&&\tilde{H}_{B}=\int d^{D-1}x\left( -A_{0}^{a}\left( \partial _{j}\pi
_{a}^{j}-f_{\;\;ac}^{b}\pi _{b}^{j}A_{j}^{c}-4g\pi ^{ij}\partial
_{i}A_{aj}\right) -2B^{0i}\partial ^{j}\pi _{ji}-\right.   \nonumber
\label{63} \\
&&\frac{1}{2}\left( \pi _{i}^{a}+4gA^{ak}\pi _{ik}\right) \left( \pi
_{a}^{i}+4gA_{aj}\pi ^{ij}\right) +\frac{1}{4}F_{ij}^{a}F_{a}^{ij}+\frac{1}{%
12}H_{ijk}H^{ijk}-  \nonumber \\
&&\pi _{ij}\pi ^{ij}+\left( \eta _{1}^{i}+2g\left(
f_{abc}A^{bi}A_{0}^{c}+\pi _{a}^{i}+4gA_{aj}\pi ^{ij}\right) \eta
_{2}^{a}\right) {\cal P}_{2i}+  \nonumber \\
&&\left. \left( \eta _{1}^{a}-f_{\;\;bc}^{a}A_{0}^{c}\eta _{2}^{b}\right) 
{\cal P}_{2a}+gf_{abc}A_{0}^{c}\eta _{2}^{a}\eta _{2}^{b}{\cal P}\right) ,
\end{eqnarray}
where 
\begin{equation}
H_{ijk}=F_{ijk}-2g\left( f_{abc}A_{i}^{a}A_{j}^{b}A_{k}^{c}+A_{\left[
i\right. }^{a}F_{a\left. jk\right] }\right) .  \label{64}
\end{equation}
From the antighost-independent terms in (\ref{62}) we observe that the
deformation of the BRST charge implies the deformed first-class constraints 
\begin{equation}
\tilde{G}_{a}^{(2)}\equiv -\left( \partial _{j}\pi
_{a}^{j}-f_{\;\;ac}^{b}\pi _{b}^{j}A_{j}^{c}-4g\pi ^{ij}\partial
_{i}A_{aj}\right) \approx 0,  \label{65}
\end{equation}
the remaining constraints in (\ref{4}--\ref{5}) being undeformed. Moreover,
the term $gf_{\;\;ab}^{c}A_{c}^{i}{\cal P}_{2i}\eta _{2}^{a}\eta _{2}^{b}$
shows that the Poisson brackets among the new constraint functions $\tilde{G}%
_{a}^{(2)}$ are also deformed like 
\begin{equation}
\left[ \tilde{G}_{a}^{(2)},\tilde{G}_{b}^{(2)}\right] =f_{\;\;ab}^{c}\left( 
\tilde{G}_{c}^{(2)}+2gA_{c}^{i}G_{i}^{(2)}\right) ,  \label{66}
\end{equation}
so the first-class constraint algebra becomes open. On the other hand, the
antighost-independent piece in (\ref{63}) 
\begin{eqnarray}
&&\tilde{H}=\int d^{D-1}x\left( -A_{0}^{a}\left( \partial _{j}\pi
_{a}^{j}-f_{\;\;ac}^{b}\pi _{b}^{j}A_{j}^{c}-4g\pi ^{ij}\partial
_{i}A_{aj}\right) -\right.   \nonumber  \label{67} \\
&&2B^{0i}\partial ^{j}\pi _{ji}-\frac{1}{2}\left( \pi _{i}^{a}+4gA^{ak}\pi
_{ik}\right) \left( \pi _{a}^{i}+4gA_{aj}\pi ^{ij}\right) -  \nonumber \\
&&\left. \pi _{ij}\pi ^{ij}+\frac{1}{4}F_{ij}^{a}F_{a}^{ij}+\frac{1}{12}%
H_{ijk}H^{ijk}\right) ,
\end{eqnarray}
is nothing but the first-class Hamiltonian of the deformed theory. The
components linear in the antighost number one antighosts from (\ref{63})
emphasize that the Poisson brackets among the new first-class Hamiltonian
and new first-class constraint functions $\tilde{G}_{a}^{(2)}$ are modified
as 
\begin{equation}
\left[ \tilde{H},\tilde{G}_{a}^{(2)}\right] =-f_{\;\;ab}^{c}A_{0}^{b}\left( 
\tilde{G}_{c}^{(2)}+2gA_{c}^{i}G_{i}^{(2)}\right) +2g\left( \pi
_{a}^{i}+4gA_{aj}\pi ^{ij}\right) G_{i}^{(2)},  \label{68}
\end{equation}
the others being kept unchanged with respect to the uncoupled model. The
resulting first-class Hamiltonian and gauge algebra describe the Yang-Mills
Chern-Simons couplings among a Yang-Mills-2-form system, known as the
Chapline-Manton model. As the first-class constraints generate
gauge transformations, from the deformations (\ref{65}--\ref{66}) we can
conclude that the added interactions involved with (\ref{67}) modify both
the gauge transformations and their gauge algebra. However, our procedure
does not affect in any way the reducibility relations of the uncoupled
theory.

The Lagrangian version of the resulting deformed model can be derived as
usually, via employing the extended and total formalisms, which then produce
nothing but the well-known Lagrangian action \cite{18}--\cite{21} 
\begin{equation}  \label{75}
\tilde S_0^L\left[ A_\mu ^a,B_{\mu \nu }\right] =\int d^Dx\left( -\frac
14F_{\mu \nu }^aF_a^{\mu \nu }-\frac 1{12}H_{\mu \nu \rho }H^{\mu \nu \rho
}\right) ,
\end{equation}
subject to the gauge transformations 
\begin{equation}  \label{76}
\delta _\epsilon A_\mu ^a=\left( D_\mu \right) _{\;\;b}^a\epsilon
^b,\;\delta _\epsilon B_{\mu \nu }=\partial _{\left[ \mu \right. }\epsilon
_{\left. \nu \right] }+2g\epsilon _a\partial _{\left[ \mu \right. }A_{\left.
\nu \right] }^a,
\end{equation}
where 
\begin{equation}  \label{77}
H_{\mu \nu \rho }=F_{\mu \nu \rho }-2g\left( f_{abc}A_\mu ^aA_\nu ^bA_\rho
^c+A_{\left[ \mu \right. }^aF_{a\left. \nu \rho \right] }\right) ,
\end{equation}
and $\left( D_\mu \right) _{\;\;b}^a=\delta _{\;\;b}^a\partial _\mu
+f_{\;\;bc}^aA_\mu ^c$ is the covariant derivative. It is precisely the
piece linear in the deformation parameter from (\ref{65}) that leads to the
second term in the Lagrangian gauge transformations of $B_{\mu \nu }$.

\section{Conclusion}

To conclude with, in this paper we have derived the consistent interactions
that can be introduced among Yang-Mills gauge fields and an abelian
two-form. Beginning with the BRST differential for the uncoupled model, we
have initially deduced the first-order deformation of the BRST charge by
expanding the co-cycles accordingly the antighost number. Subsequently, we
have shown that this deformation is consistent also at higher-orders. In the
next step we have determined a deformed BRST-invariant Hamiltonian, that is
quadratic in the deformation parameter. In this manner, we have generated
precisely the combined Yang-Mills-two-form system coupled through Yang-Mills
Chern-Simons term. The added interactions deform both the gauge
transformations and gauge algebra, but not the reducibility relations.

\section*{Acknowledgment}

Two of the authors (C.B. and S.O.S.) acknowledge financial support from a
Romanian National Council for Academic Scientific Research (CNCSIS) grant.

\end{document}